\documentstyle[sprocl,epsf]{article}

\def\gsim{\ \rlap{\raise 3pt \hbox{$>$}}{\lower 3pt \hbox{$\sim$}}\ }
\def\lsim{\ \rlap{\raise 3pt \hbox{$<$}}{\lower 3pt \hbox{$\sim$}}\ }

\begin{document}

\begin{titlepage}

\begin{flushright}
CERN-TH/98-179\\
hep-ph/9806206
\end{flushright}

\vspace{1.5cm}

\begin{center}
\Large\bf 
Looking for the Unexpected:\\ 
Direct CP Violation in \boldmath$B\to X_s\gamma$\unboldmath\ Decays
\end{center}

\vspace{1.2cm}

\begin{center}
Matthias Neubert\\
{\sl Theory Division, CERN, CH-1211 Geneva 23, Switzerland}
\end{center}

\vspace{1.3cm}

\begin{center}
{\bf Abstract:}\\[0.3cm]
\parbox{11cm}{
The observation of a sizable direct CP asymmetry in the inclusive
decays $B\to X_s\gamma$ would be a clean signal of New Physics. In the
Standard Model, this asymmetry is below 1\% in magnitude. In extensions
of the Standard Model with new CP-violating couplings, large
asymmetries are possible without conflicting with the experimental
value of the $B\to X_s\gamma$ branching ratio. In particular, large
asymmetries arise naturally in models with enhanced chromo-magnetic
dipole transitions. Some generic examples of such models are explored
and their implications for the semileptonic branching ratio and charm
yield in $B$ decays discussed.}
\end{center}

\vspace{1cm}

\begin{center}
{\sl To appear in the Proceedings of the\\
Third Workshop on Continuous Advances in QCD\\
Minneapolis, Minnesota, 16--19 April 1998}
\end{center}

\vspace{1.5cm}

\vfil
\noindent
CERN-TH/98-179\\
June 1998

\end{titlepage}

\thispagestyle{empty}
\vbox{}
\newpage

\setcounter{page}{1}


\title{LOOKING FOR THE UNEXPECTED:\\ 
DIRECT CP VIOLATION IN \boldmath$B\to X_s\gamma$\unboldmath\ DECAYS}

\author{M. NEUBERT}

\address{Theory Division, CERN, CH-1211 Geneva 23, Switzerland\\
E-mail: Matthias.Neubert@cern.ch} 

\maketitle\abstracts{
The observation of a sizable direct CP asymmetry in the inclusive
decays $B\to X_s\gamma$ would be a clean signal of New Physics. In the
Standard Model, this asymmetry is below 1\% in magnitude. In extensions
of the Standard Model with new CP-violating couplings, large
asymmetries are possible without conflicting with the experimental
value of the $B\to X_s\gamma$ branching ratio. In particular, large
asymmetries arise naturally in models with enhanced chromo-magnetic
dipole transitions. Some generic examples of such models are explored
and their implications for the semileptonic branching ratio and charm
yield in $B$ decays discussed.}

\section{Introduction}

Studies of rare decays of $B$ mesons have the potential to uncover the
origin of CP violation and provide hints to physics beyond the Standard
Model of strong and electroweak interactions. The measurements of
several CP asymmetries will make it possible to test whether the CKM
paradigm is correct, or whether additional sources of CP violation are
required. In order to achieve this goal, it is necessary that the
theoretical calculations of CP-violating observables in terms of
Standard Model parameters are, to a large extent, free of hadronic
uncertainties. This can be achieved, e.g., by measuring time-dependent
asymmetries in the decays of neutral $B$ mesons into particular CP
eigenstates. In many other cases, however, the theoretical predictions
for direct CP violation in exclusive $B$ decays are obscured by large
strong-interaction effects \cite{Blok96}$^-$\cite{At97}.

Inclusive decay rates of $B$ mesons, on the other hand, can be reliably
calculated in QCD using the operator product expansion. Up to small
bound-state corrections these rates agree with the parton model
predictions for the underlying decays of the $b$ quark. The
disadvantage that the sum over many final states partially dilutes the
CP asymmetries in inclusive decays is compensated by the fact that,
because of the short-distance nature of these processes, the strong
phases are calculable using quark--hadron duality. In this talk, I
report on a study \cite{Alex1} of direct CP violation in the rare
radiative decays $B\to X_s\gamma$, both in the Standard Model and
beyond. These decays have already been observed experimentally, and
copious data samples will be collected at the $B$ factories. The
theoretical analysis relies only on the weak assumption of global
quark--hadron duality, and the leading nonperturbative corrections are
well understood.

We perform a model-independent analysis of CP-violating effects in
$B\to X_s\gamma$ decays in terms of the effective Wilson coefficients
$C_7\equiv C_7^{\rm eff}(m_b)$ and $C_8\equiv C_8^{\rm eff}(m_b)$
multiplying the (chromo-) magnetic dipole operators $O_7=e\,m_b\,\bar
s_L\sigma_{\mu\nu} F^{\mu\nu} b_R$ and $O_8=g_s m_b\,\bar
s_L\sigma_{\mu\nu} G^{\mu\nu} b_R$ in the effective weak Hamiltonian.
We allow for generic New Physics contributions to these coefficients.
Several extensions of the Standard Model in which such contributions
arise have been explored, e.g., in Refs.~7--10.
We find that in the Standard Model the direct CP asymmetry in $B\to
X_s\gamma$ decays is very small (below 1\% in magnitude) because of a
combination of CKM and GIM suppression, both of which can be lifted in
New Physics scenarios with additional contributions to the dipole
operators containing new weak phases. We thus propose a measurement of
the inclusive CP asymmetry in radiative $B$ decays as a clean and
sensitive probe of New Physics. Studies of direct CP violation in the
inclusive decays $B\to X_s\gamma$ have been performed previously by
several authors, both in the Standard Model \cite{Soares} and in
certain extensions of it \cite{Wolf,Asat}. In all cases rather small
asymmetries were obtained. We generalize and extend these analyses in
various ways. Besides including some contributions neglected in
previous works, we investigate a class of New Physics models with
enhanced chromo-magnetic dipole contributions, in which large CP
asymmetries of order 10--50\% are possible and even natural. We also
employ a full next-to-leading order analysis of the CP-averaged $B\to
X_s\gamma$ branching ratio in order to derive constraints on the
parameter space of the New Physics models considered.

\boldmath
\section{Direct CP violation in radiative $B$ decays}
\unboldmath
\label{sec:ACP}

The starting point in the calculation of the inclusive $B\to X_s\gamma$
decay rate is provided by the effective weak Hamiltonian renormalized
at the scale $\mu=m_b$. Direct CP violation in these decays may arise
from the interference of non-trivial weak phases, contained in CKM
parameters or in possible New Physics contributions to the Wilson
coefficient functions, with strong phases provided by the imaginary
parts of the matrix elements of the operators in the effective
Hamiltonian \cite{Band}. These imaginary parts first arise at
$O(\alpha_s)$ from loop diagrams containing charm quarks, light quarks
or gluons. Using the formulae of Greub et al.\ for these contributions
\cite{Greub}, we calculate at next-to-leading order the difference
$\Delta\Gamma=\Gamma(\bar B\to X_s\gamma)-\Gamma(B\to X_{\bar
s}\gamma)$ of the CP-conjugate, inclusive decay rates. The
contributions to $\Delta\Gamma$ from virtual corrections arise from
interference of the one-loop diagrams with insertions of the operators
$O_2$ and $O_8$ with the tree-level diagram containing $O_7$. Here
$O_2=\bar s_L\gamma_\mu q_L\,\bar q_L\gamma^\mu b_L$ with $q=c,u$ are
the usual current--current operators in the effective Hamiltonian.
There are also contributions to $\Delta\Gamma$ from gluon
bremsstrahlung diagrams with a charm-quark loop. They can interfere
with the tree-level diagrams for $b\to s\gamma g$ containing an
insertion of $O_7$ or $O_8$. Contrary to the virtual corrections, for
which in the parton model the photon energy is fixed to its maximum
value, the gluon bremsstrahlung diagrams lead to a non-trivial photon
spectrum, and so the results depend on the experimental lower cutoff on
the photon energy. We define a quantity $\delta$ by the requirement
that $E_\gamma > (1-\delta) E_\gamma^{\rm max}$. Combining the two
contributions and dividing the result by the leading-order expression
for twice the CP-averaged inclusive decay rate, we find for the CP
asymmetry
\begin{eqnarray}
   A_{\rm CP}^{b\to s\gamma}(\delta)
   &=& \frac{\Gamma(\bar B\to X_s\gamma)-\Gamma(B\to X_{\bar s}\gamma)}
            {\Gamma(\bar B\to X_s\gamma)+\Gamma(B\to X_{\bar s}\gamma)}
    \Bigg|_{E_\gamma>(1-\delta) E_\gamma^{\rm max}} \nonumber\\
   &=& \frac{\alpha_s(m_b)}{|C_7|^2}\,\Bigg\{
    \frac{40}{81}\,\mbox{Im}[C_2 C_7^*]
    - \frac{8z}{9}\,\Big[ v(z) + b(z,\delta) \Big]\,
    \mbox{Im}[(1+\epsilon_s) C_2 C_7^*] \nonumber\\
   &&\hspace{1.35cm}
    \mbox{}- \frac 49\,\mbox{Im}[C_8 C_7^*]
    + \frac{8z}{27}\,b(z,\delta)\,\mbox{Im}[(1+\epsilon_s) C_2 C_8^*]
    \Bigg\} \,,
\label{ACP}
\end{eqnarray}
where $z=(m_c/m_b)^2$, and the explicit expressions for the functions
$g(z)$ and $b(z,\delta)$ can be found in Ref.~6.
The
quantity $\epsilon_s$ is a ratio of CKM matrix elements given by
\begin{equation}
   \epsilon_s = \frac{V_{us}^* V_{ub}}{V_{ts}^* V_{tb}}
   \approx \lambda^2 (i\eta-\rho) = O(10^{-2}) \,,
\end{equation}
where $\lambda=\sin\theta_{\rm C}\approx 0.22$ and $\rho,\eta=O(1)$ are
the Wolfenstein parameters. An estimate of the $C_2$--$C_7$
interference term in (\ref{ACP}) was obtained previously by Soares
\cite{Soares}, who neglects the contribution of $b(z,\delta)$ and uses
an approximation for the function $v(z)$. The relevance of the
$C_8$--$C_7$ interference term for two-Higgs-doublet models, and for
left--right symmetric extensions of the Standard Model, has been
explored in Refs.~12,13.

In the Standard Model, the Wilson coefficients take the real values
$C_2\approx 1.11$, $C_7\approx -0.31$ and $C_8\approx -0.15$. The
imaginary part of the small quantity $\epsilon_s$ is thus the only
source of CP violation. Note that all terms involving this quantity are
GIM suppressed by a power of the small ratio $z=(m_c/m_b)^2$,
reflecting the fact that there is no non-trivial weak phase difference
in the limit where $m_c=m_u=0$. Hence, the Standard Model prediction
for the CP asymmetry is suppressed by three small factors:
$\alpha_s(m_b)$ arising from the strong phases, $\sin^2\!\theta_{\rm
C}$ reflecting the CKM suppression, and $(m_c/m_b)^2$ resulting from
the GIM suppression. The numerical result for the asymmetry depends on
the values of the strong coupling constant and the ratio of the
heavy-quark pole masses, for which we take $\alpha_s(m_b)\approx 0.214$
(corresponding to $\alpha_s(m_Z)=0.118$ and two-loop evolution down to
the scale $m_b=4.8$\,GeV) and $\sqrt z=m_c/m_b=0.29$. This yields
$A_{\rm CP,SM}^{b\to s\gamma} \approx (1.5\mbox{--}1.6)\%\,\eta$
depending on the value of $\delta$. With $\eta\approx 0.2$--0.4 as
suggested by phenomenological analyses, we find a tiny asymmetry of
about 0.5\%, in agreement with the estimate obtained in
Ref.~11.
Expression (\ref{ACP}) applies also to the
decays $B\to X_d\,\gamma$, the only difference being that in this case
the quantity $\epsilon_s$ must be replaced with the corresponding
quantity $\epsilon_d = (V_{ud}^* V_{ub})/(V_{td}^* V_{tb}) \approx
(\rho-i\eta)/(1-\rho+i\eta) = O(1)$. Therefore, in the Standard Model
the CP asymmetry in $B\to X_d\,\gamma$ decays is larger by a factor of
about $-20$ than that in $B\to X_s\gamma$ decays. However,
experimentally it is difficult to distinguish between $B\to X_s\gamma$
and $B\to X_d\,\gamma$ decays. If only their sum is measured, the CP
asymmetry vanishes by CKM unitarity \cite{Soares}.

\begin{table}
\caption{Values of the coefficients $a_{ij}$ in \%, without (left) and
with (right) Fermi motion effects included}
\vspace{0.2cm}
\begin{center}
\begin{tabular}{|cc|ccc|ccc|}
\hline
$\delta$ & $E_\gamma^{\rm min}~[{\rm GeV}]$ & $a_{27}$ & $a_{87}$
 & $a_{28}$ & $a_{27}$ & $a_{87}$ & $a_{28}$ \\
 & & \multicolumn{3}{c|}{(parton model)}
 & \multicolumn{3}{c|}{(with Fermi motion)} \\
\hline\hline
1.00 & 0.00 & 1.06 & $-9.52$ & 0.16 & 1.06 & $-9.52$ & 0.16 \\
0.30 & 1.85 & 1.17 & $-9.52$ & 0.12 & 1.23 & $-9.52$ & 0.10 \\
0.15 & 2.24 & 1.31 & $-9.52$ & 0.07 & 1.40 & $-9.52$ & 0.04 \\
\hline
\end{tabular}
\end{center}
\label{tab:aij}
\end{table}

{}From (\ref{ACP}) it is apparent that two of the suppression factors
operative in the Standard Model, $z$ and $\lambda^2$, can be avoided in
models where the effective Wilson coefficients $C_7$ and $C_8$ receive
additional contributions involving non-trivial weak phases. Much larger
CP asymmetries then become possible. In order to investigate such
models, we may to good approximation neglect the small quantity
$\epsilon_s$ and write
\begin{equation}
   A_{\rm CP}^{b\to s\gamma}(\delta)
   = a_{27}(\delta)\,\mbox{Im}\!\left[ \frac{C_2}{C_7} \right]
   + a_{87}\,\mbox{Im}\!\left[ \frac{C_8}{C_7} \right]
   + a_{28}(\delta)\,\frac{\mbox{Im}[C_2 C_8^*]}{|C_7|^2} \,.
\label{3terms}
\end{equation}
The values of the coefficients $a_{ij}$ are shown in the left portion
of Table~\ref{tab:aij} for three choices of the cutoff on the photon
energy: $\delta=1$ corresponding to the (unrealistic) case of a fully
inclusive measurement, $\delta=0.3$ corresponding to a restriction to
the part of the spectrum above $1.85$\,GeV, and $\delta=0.15$
corresponding to a cutoff that removes almost all of the background
from $B$ decays into charmed hadrons. In practice, a restriction to the
high-energy part of the photon spectrum is required for experimental
reasons. Note, however, that the result for the CP asymmetry is not
very sensitive to the choice of the photon-energy cutoff. Whereas the
third term in (\ref{3terms}) is generally very small, the first two
terms can give rise to sizable effects. Assume, e.g., that there is a
New Physics contribution to $C_7$ of similar magnitude as the Standard
Model contribution (so as not to spoil the prediction for the $B\to
X_s\gamma$ branching ratio) but with a non-trivial weak phase. Then the
first term in (\ref{3terms}) may give a contribution of up to about 5\%
in magnitude. Similarly, if there are New Physics contributions to
$C_7$ and $C_8$ such that the ratio $C_8/C_7$ has a non-trivial weak
phase, the second term may give a contribution of up to about
$10\%\times|C_8/C_7|$. In models with a strong enhancement of $|C_8|$
with respect to its Standard Model value, there is thus the possibility
of generating a large direct CP asymmetry in $B\to X_s\gamma$ decays.

In our discussion so far we have neglected nonperturbative power
corrections to the inclusive decay rates. Their impact on the rate
ratio defining the CP asymmetry is very small, since most of the
corrections cancel between the numerator and the denominator.
Potentially the most important bound-state effect is the Fermi motion
of the $b$ quark inside the $B$ meson, which determines the shape of
the photon energy spectrum in the endpoint region. This effect is
included in the heavy-quark expansion by resumming an infinite set of
leading-twist corrections into a ``shape function'', which governs the
momentum distribution of the heavy quark inside the meson
\cite{shape,Dike95}. The physical decay distributions are obtained from
a convolution of parton model spectra with this function. In the
process, phase-space boundaries defined by parton kinematics are
transformed into the proper physical boundaries defined by hadron
kinematics. Details of the implementation of this effect can be found
in Refs.~6,18,
where a simple ansatz for the shape
function is employed. The right portion of Table~\ref{tab:aij} shows
the values of the coefficients $a_{ij}(\delta)$ corrected for Fermi
motion. The largest coefficient, $a_{87}$, is not affected, and the
impact on the other two coefficients is rather mild. As a consequence,
the predictions for the CP asymmetry are quite insensitive to
bound-state effects, even if a restriction on the high-energy part of
the photon spectrum is imposed.

In the next section we explore the structure of New Physics models with
a potentially large inclusive CP asymmetry. A non-trivial constraint on
such models is that they must yield an acceptable result for the total,
CP-averaged $B\to X_s\gamma$ branching ratio, which has been measured
experimentally. Taking a weighed average of the results reported by the
CLEO and ALEPH Collaborations \cite{CLEO,ALEPH} gives $\mbox{B}(B\to
X_s\gamma)=(2.5\pm 0.6)\times 10^{-4}$. The complete theoretical
prediction for the $B\to X_s\gamma$ branching ratio at next-to-leading
order has been presented for the first time by Chetyrkin et al.\
\cite{Chet}, and subsequently has been discussed by several authors
\cite{Buras}$^-$\cite{newGreub}. It depends on the Wilson coefficients
$C_2$, $C_7$ and $C_8$ through the combinations $\mbox{Re}[C_i C_j^*]$.
Recently, we have extended these analyses in several aspects, including
a discussion of Fermi motion effects and a conservative analysis of
truncation errors \cite{newpaper}. In contrast to the case of the CP
asymmetry, Fermi motion effects are very important when comparing
experimental data for the $B\to X_s\gamma$ branching ratio with
theoretical predictions. With our choice of parameters, we obtain for
the total branching ratio in the Standard Model $\mbox{B}(B\to
X_s\gamma)=(3.3\pm 0.3)\times 10^{-4}$, which is consistent with the
experimental findings.

\section{CP asymmetry beyond the Standard Model}

In order to explore the implications of various New Physics scenarios
for the CP asymmetry and branching ratio in $B\to X_s\gamma$ decays it
is useful to express the Wilson coefficients $C_7=C_7^{\rm eff}(m_b)$
and $C_8=C_8^{\rm eff}(m_b)$, which are defined at the scale $m_b$, in
terms of their values at the high scale $m_W$, using the
renormalization group. This yields
\begin{eqnarray}
   C_7 &=& \eta^\frac{16}{23}\,C_7(m_W) + \frac 83 \left(
    \eta^\frac{14}{23} - \eta^\frac{16}{23} \right) C_8(m_W)
    + \sum_{i=1}^8\,h_i\,\eta^{a_i} \,, \nonumber\\
   C_8 &=& \eta^\frac{14}{23}\,C_8(m_W)
    + \sum_{i=1}^8\,\bar h_i\,\eta^{a_i} \,,
\label{evol}
\end{eqnarray}
where $\eta=\alpha_s(m_W)/\alpha_s(m_b)\approx 0.56$, and $h_i$, $\bar
h_i$ and $a_i$ are known numerical coefficients \cite{Guidoetal,Poko}.
For the Wilson coefficients at the scale $m_W$, we write $C_{7,8}(m_W)
= C_{7,8}^{\rm SM}(m_W)+ C_{7,8}^{\rm new}(m_W)$. The first term
corresponds to the Standard Model contributions \cite{Gr90}, which are
functions of the mass ratio $x_t=(m_t/m_W)^2$. Numerically, one obtains
\begin{eqnarray}
   C_7 &\approx& -0.31 + 0.67\,C_7^{\rm new}(m_W)
    + 0.09\,C_8^{\rm new}(m_W) \,, \nonumber\\
   C_8 &\approx& -0.15 + 0.70\,C_8^{\rm new}(m_W) \,.
\label{C7C8}
\end{eqnarray}
We choose to parametrize our results in terms of the magnitude and
phase of one of the New Physics contributions, $C_8^{\rm
new}(m_W)\equiv K_8\,e^{i\gamma_8}$ or $C_7^{\rm new}(m_W)\equiv-
K_7\,e^{i\gamma_7}$, as well as the ratio
\begin{equation}
   \xi = \frac{C_7^{\rm new}(m_W)}{Q_d\,C_8^{\rm new}(m_W)} \,,
\label{xidef}
\end{equation}
where $Q_d=-\frac 13$. A given New Physics scenario predicts these
quantities at some large scale $M$. Using the renormalization group, it
is then possible to evolve these predictions down to the scale $m_W$.
Typically, $\xi\equiv\xi(m_W)$ tends to be smaller than $\xi(M)$ by an
amount of order $-0.1$ to $-0.3$, depending on how close the New
Physics is to the electroweak scale. We restrict ourselves to cases
where the parameter $\xi$ in (\ref{xidef}) is real; otherwise there
would be even more potential for CP violation. This happens if there is
a single dominant New Physics contribution, such as the virtual
exchange of a new heavy particle, contributing to both the magnetic and
the chromo-magnetic dipole operators.

\begin{table}
\caption{Ranges of $\xi(M)$ for various New Physics contributions to
$C_7$ and $C_8$, characterized by the particles in penguin diagrams}
\vspace{0.2cm}
\begin{center}
\begin{tabular}{|lc|}
\hline
Class-1 models & $\xi(M)$ \\
\hline\hline
neutral scalar--vectorlike quark & 1 \\
gluino--squark ($m_{\tilde g} < 1.37 m_{\tilde q}$)
 & $-(0.13\mbox{--}1)$ \\
techniscalar & $\approx-0.5$ \\
\hline\hline
Class-2 models & $\xi(M)$ \\
\hline\hline
scalar diquark--top & 4.8--8.3 \\
gluino--squark ($m_{\tilde g} > 1.37 m_{\tilde q}$)
 & $-(1\mbox{--}2.9)$ \\
charged Higgs--top & $-(2.4\mbox{--}3.8)$ \\
left--right $W$--top & $\approx -6.7$ \\
Higgsino--stop & $-(2.6\mbox{--}24)$ \\
\hline
\end{tabular}
\end{center}
\label{tab:xi}
\end{table}

\begin{figure}
\epsfxsize=11.9cm
\centerline{\epsffile{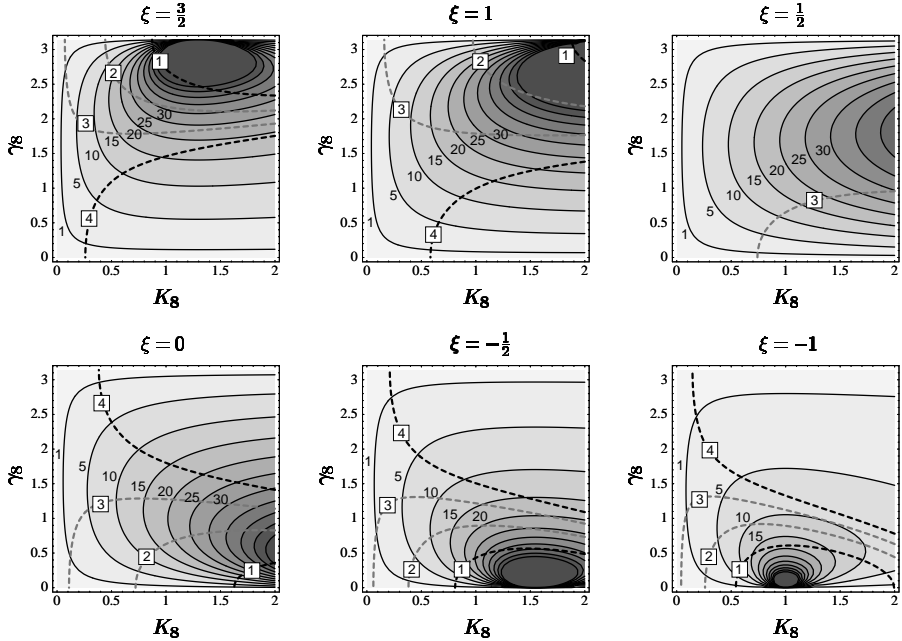}}
\caption{Contour plots for the CP asymmetry $A_{\rm CP}^{b\to s\gamma}$
for various class-1 models. We show contours only until values $A_{\rm
CP}=50\%$; for such large values, the theoretical expression for the CP
asymmetry in (\protect\ref{3terms}) would have to be extended to higher
orders to get a reliable result.}
\label{fig:models1}
\end{figure}

Ranges of $\xi(M)$ for several illustrative New Physics scenarios are
collected in Table~\ref{tab:xi}. For a detailed discussion of the model
parameters which lead to the $\xi$ values quoted in the table the
reader is referred to Ref.~6.
Our aim is not to carry out
a detailed study of each model, but to give an idea of the sizable
variation that is possible in $\xi$. It is instructive to distinguish
two classes of models: those with moderate (class-1) and those with
large (class-2) values of $|\xi|$. It follows from (\ref{C7C8}) that
for small positive values of $\xi$ it is possible to have large complex
contributions to $C_8$ without affecting too much the magnitude and
phase of $C_7$, since
\begin{equation}
   \frac{C_8}{C_7}\approx\frac{0.70 K_8\,e^{i\gamma_8}-0.15}
    {(0.09-0.22\xi) K_8\,e^{i\gamma_8}-0.31} \,.
\label{C7C8rat}
\end{equation}
This is also true for small negative values of $\xi$, albeit over a
smaller region of parameter space. New Physics scenarios that have this
property belong to class-1 and have been explored in
Ref.~7.
They allow for large CP asymmetries resulting from
the $C_7$--$C_8$ interference term in (\ref{3terms}).
Figure~\ref{fig:models1} shows contour plots for the CP asymmetry in
the $(K_8,\gamma_8)$ plane for six different choices of $\xi$ between
$\frac32$ and $-1$, assuming a cutoff $E_\gamma>1.85$\,GeV on the
photon energy. We repeat that the results for the CP asymmetry depend
very little on the choice of the cutoff. For each value of $\xi$, the
plots cover the region $0\le K_8\le 2$ and $0\le\gamma_8\le\pi$
(changing the sign of $\gamma_8$ would only change the sign of the CP
asymmetry). The contour lines refer to values of the asymmetry of 1\%,
5\%, 10\%, 15\% etc. The thick dashed lines indicate contours where the
$B\to X_s\gamma$ branching ratio takes values between $1\times 10^{-4}$
and $4\times 10^{-4}$, as indicated by the numbers inside the squares.
The Standard Model prediction with this choice of the photon-energy
cutoff is about $3\times 10^{-4}$. The main conclusion to be drawn from
Figure~\ref{fig:models1} is that in class-1 scenarios there exists
great potential for sizable CP asymmetries in a large region of
parameter space, whereas any point to the right of the 1\% contour for
$A_{\rm CP}^{b\to s\gamma}$ cannot be accommodated by the Standard
Model. Note that quite generally the regions of parameter space that
yield large values for the CP asymmetries are not excluded by the
experimental constraint on the CP-averaged branching ratio. To have
large CP asymmetries the products $C_i C_j^*$ are required to have
large imaginary parts [cf.\ (\ref{3terms})], whereas the total
branching ratio is sensitive to the real parts of these quantities.

\begin{figure}
\epsfxsize=11.9cm
\centerline{\epsffile{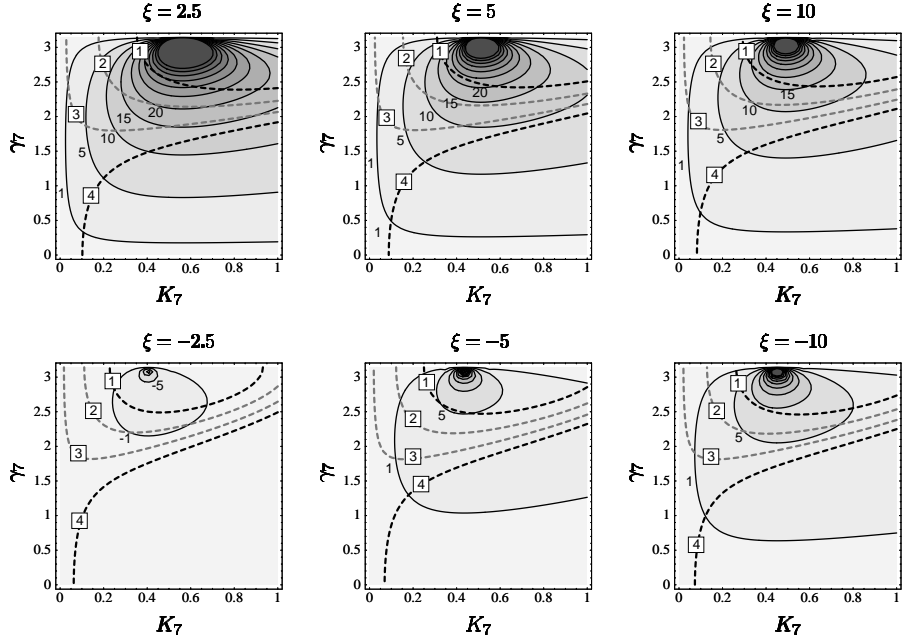}}
\caption{Contour plots for the CP asymmetry $A_{\rm CP}^{b\to s\gamma}$
for various class-2 models}
\label{fig:models2}
\end{figure}

There are also scenarios in which the parameter $\xi$ takes on larger
negative or positive values. In such cases, it is not possible to
increase the magnitude of $C_8$ much over its Standard Model value, and
the only way to get large CP asymmetries from the $C_7$--$C_8$ or
$C_7$--$C_2$ interference terms in (\ref{3terms}) is to have $C_7$
tuned to be very small; however, this possibility is constrained by the
fact that the total $B\to X_s\gamma$ branching ratio must be of an
acceptable magnitude. That this condition starts to become a limiting
factor is already seen in the plots corresponding to $\xi=-\frac12$ and
$-1$ in Figure~\ref{fig:models1}. For even larger values of $|\xi|$,
the $C_7$--$C_8$ interference term becomes ineffective, because the
weak phase tends to cancel in the ratio $C_8/C_7$. Then the
$C_2$--$C_7$ interference term becomes the main source of CP violation;
however, as discussed in Section~\ref{sec:ACP}, it cannot lead to
asymmetries exceeding a level of about 5\% without violating the
constraint that the $B\to X_s\gamma$ branching ratio not be too small.
Models of this type belong to the class-2 category. For a graphical
analysis of class-2 models it is convenient to choose the magnitude and
phase of the New Physics contribution $C_7^{\rm new}(m_W)\equiv
-K_7\,e^{i\gamma_7}$ as parameters, rather than $K_8$ and $\gamma_8$.
The reason is that for large $|\xi|$ it becomes increasingly unlikely
that $C_8^{\rm new}(m_W)$ will be large. The resulting plots are given
in Figure \ref{fig:models2}. The branching-ratio constraint allows
larger values of $C_8$ for positive $\xi$, which explains why larger
asymmetries are attainable in this case. For example, for $\xi\approx
5$, which can be obtained from scalar diquark--top penguins,
asymmetries of 5--20\% are seen to be consistent with the $B\to
X_s\gamma$ bound. On the other hand, for $\xi\approx-(2.5\mbox{--}5)$,
which includes the multi-Higgs-doublet models, CP asymmetries of only a
few percent are attainable, in agreement with the findings of previous
authors \cite{Wolf,Asat,newGreub}. The same is true for the left--right
symmetric $W$--top penguin, particularly if one takes into account that
$K_7\lsim 0.2$ if $m_{W_R}>1$\,TeV.

\begin{figure}
\epsfxsize=9cm
\centerline{\epsffile{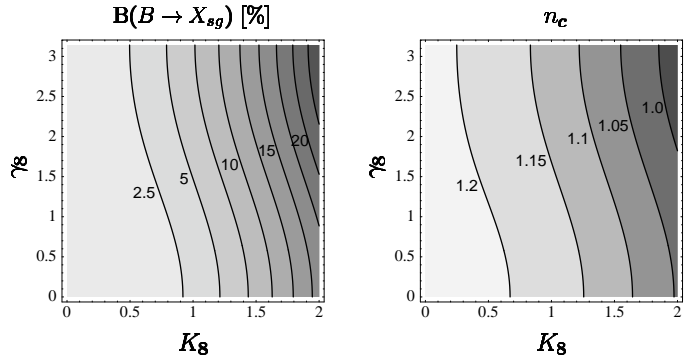}}
\caption{Contour plots for the $B\to X_{sg}$ branching ratio (left) and
for the charm yield $n_c$ in $B$ decays (right). There is an overall
theoretical uncertainty of 6\% on the values of $n_c$.}
\label{fig:nc}
\end{figure}

The class-1 New Physics scenarios explored in Figure~\ref{fig:models1}
have the attractive feature of a possible large enhancement of the
magnitude of the Wilson coefficient $C_8$. This could have important
implications for the phenomenology of the semileptonic branching ratio
and charm yield in $B$ decays, through enhanced production of charmless
hadronic final states induced by the $b\to s g$
transition~\cite{Kaga,CGG95,hou}. At $O(\alpha_s)$, the theoretical
expression for the $B\to X_{sg}$ branching ratio is proportional to
$|C_8|^2$. The left-hand plot in Figure~\ref{fig:nc} shows contours for
this branching ratio in the $(K_8,\gamma_8)$ plane. In the Standard
Model, $\mbox{B}(B\to X_{sg})\approx 0.2\%$ is very small; however, in
scenarios with $|C_8|=O(1)$ sizable values of order 10\% for this
branching ratio are possible, which simultaneously lowers the
theoretical predictions for the semileptonic branching ratio and the
charm production rate $n_c$ by a factor of $[1+\mbox{B}(B\to
X_{sg})]^{-1}$. The most recent value of $n_c$ reported by the CLEO
Collaboration is \cite{Drell} $1.12\pm 0.05$. Although the systematic
errors in this measurement are large, the result favours values of
$\mbox{B}(B\to X_{sg})$ of order 10\%. This is apparent from the
right-hand plot in Figure~\ref{fig:nc}, which shows the central
theoretical prediction for $n_c$ as a function of $K_8$ and $\gamma_8$.
(There is an overall theoretical uncertainty in the value of $n_c$ of
about 6\% resulting from the dependence on quark masses and the
renormalization scale \cite{NeSa}.) The theoretical prediction for the
semileptonic branching ratio would have the same dependence on $K_8$
and $\gamma_8$, with the normalization ${\rm B_{SL}}=(12\pm 1)\%$ fixed
at $K_8=0$. A large value of $\mbox{B}(B\to X_{sg})$ could also help in
understanding the $\eta'$ yields in charmless $B$ decays
\cite{Houeta,Petrov}. For completeness, we note that the CLEO
Collaboration has recently presented a preliminary upper limit on
$\mbox{B}(B\to X_{sg})$ of 6.8\% (90\% CL) \cite{Thorn}. It is
therefore worth noting that large CP asymmetries of order 10--20\% can
also be attained at smaller $B\to X_{sg}$ branching ratios of a few
percent, which would nevertheless represent a marked departure from the
Standard Model prediction.

\section{Conclusions}

I have reported on a study of direct CP violation in the inclusive,
radiative decays $B\to X_s\gamma$. From a theoretical point of view,
inclusive decay rates entail the advantage of being calculable in QCD,
so that a reliable prediction for the CP asymmetry can be confronted
with data. From a practical point of view, it is encouraging that $B\to
X_s\gamma$ decays have already been observed experimentally, and
high-statistics measurements will be possible in the near future. We
find that in the Standard Model the CP asymmetry in $B\to X_s\gamma$
decays is strongly suppressed by three small parameters:
$\alpha_s(m_b)$ arising from the necessity of having strong phases,
$\sin^2\!\theta_{\rm C}\approx 5\%$ reflecting a CKM suppression, and
$(m_c/m_b)^2\approx 8\%$ resulting from a GIM suppression. As a result,
the asymmetry is only of order 1\% in magnitude -- a conclusion that
cannot be significantly modified by long-distance contributions. We
have argued that the latter two suppression factors are inoperative in
extensions of the Standard Model for which the effective Wilson
coefficients $C_7$ and $C_8$ receive additional contributions involving
non-trivial weak phases. Much larger CP asymmetries are therefore
possible in such cases.

A model-independent analysis of New Physics scenarios in terms of the
magnitudes and phases of the Wilson coefficients $C_7$ and $C_8$ shows
that, indeed, sizable CP asymmetries are predicted in large regions of
parameter space. In particular, asymmetries of 10--50\% are possible in
models which allow for a strong enhancement of the coefficient of the
chromo-magnetic dipole operator. They are, in fact, quite natural
unless there is a symmetry that forbids new weak phases from entering
the Wilson coefficients. Quite generally, having a large CP asymmetry
is not in conflict with the observed value for the CP-averaged $B\to
X_s\gamma$ branching ratio. On the contrary, it may even help to lower
the theoretical prediction for this quantity, and likewise for the
semileptonic branching ratio and charm multiplicity in $B$ decays,
thereby bringing these three observables closer to their experimental
values.

The fact that a large inclusive CP asymmetry in $B\to X_s\gamma$ decays
is possible in many generic extensions of the Standard Model, and in a
large region of parameter space, offers the possibility of looking for
a signature of New Physics in these decays using data sets that will
become available during the first period of operation of the $B$
factories. A negative result of such a study would impose constraints
on many New Physics scenarios. A positive signal, on the other hand,
would provide interesting clues about the nature of physics beyond the
Standard Model. In particular, a CP asymmetry exceeding the level of
10\% would be a strong hint towards enhanced chromo-magnetic dipole
transitions caused by some new flavour physics.

We have restricted our analysis to the case of inclusive radiative
decays since they entail the advantage of being very clean, in the
sense that the strong-interaction phases relevant for direct CP
violation can be reliably calculated. However, if there is New Physics
that induces a large inclusive CP asymmetry in $B\to X_s\gamma$ decays
it will inevitably also lead to sizable asymmetries in some related
processes. In particular, since we found that the inclusive CP
asymmetry remains almost unaffected if a cut on the high-energy part of
the photon energy spectrum is imposed, we expect that a large asymmetry
will persist in the exclusive decay mode $B\to K^*\gamma$, even though
a reliable theoretical analysis would be much more difficult because of
the necessity of calculating final-state rescattering phases
\cite{GSW95}. Still, it would be worthwhile searching for a large CP
asymmetry in this channel.

\section*{Acknowledgments}

The work reported here has been done in a most pleasant collaboration
with Alex Kagan, which is gratefully acknowledged.

\end{document}